\documentclass{jpaper}

\usepackage[nocompress]{cite}
\usepackage{algorithmic}
\usepackage{array}

\makeatletter
\let\MYcaption\@makecaption
\makeatother

\usepackage[font=footnotesize]{subcaption}

\makeatletter
\let\@makecaption\MYcaption
\makeatother

\usepackage{fixltx2e}
\usepackage{dblfloatfix}
\usepackage[nolessnomore, italic]{mathastext}
\usepackage[T1]{fontenc}
\usepackage[usenames,dvipsnames,svgnames,table]{xcolor}
\usepackage[normalem]{ulem}
\usepackage{enumitem}
\usepackage{setspace}
\usepackage{indentfirst}
\usepackage{footmisc}
\usepackage{fancyhdr}
\usepackage{authblk}
\usepackage{url}
\usepackage[binary-units=true]{siunitx}
\usepackage[us,12hr]{datetime}
\usepackage[keeplastbox]{flushend}
\usepackage[hidelinks]{hyperref}

\usepackage{xspace}

\newif\ifcameraready
\camerareadytrue

\newcommand{\versionnum}[0]{3.4}

\ifcameraready
\else
\fi

\ifcameraready
  \newcommand{\todo}[1][]{}
  \newcommand{\ch}[0]{}
\else
  \newcommand{\todo}[1][]{\textbf{\fcolorbox{black}{red}{\color{white}{TODO}}} \underline{$\overline{\hbox{\emph{#1}}}$}}
  \newcommand{\ch}[1]{{\textcolor{BrickRed}{#1}}}
\fi

\newcommand{\cmdact}{{\small{\texttt{ACTIVATE}}}\xspace}
\newcommand{\cmdrd}{{\small{\texttt{READ}}}\xspace}
\newcommand{\cmdwr}{{\small{\texttt{WRITE}}}\xspace}

\newcommand{\cmdtr}{{\small{\texttt{TRANSFER}}}\xspace}

\newcommand{\mcpy}{\mbox{\texttt{memcopy}}\xspace}
\newcommand{\minit}{\mbox{\texttt{meminit}}\xspace}

\newcommand{\fork}{\texttt{fork}\xspace}

\date{}

\begin{document}

\title{RowClone: Accelerating Data Movement \ch{and Initialization} Using DRAM}

\author{
  {Vivek Seshadri$^{1,2}$\qquad}
  Yoongu Kim$^{2}$\qquad
  Chris Fallin$^{2}$\qquad
  Donghyuk Lee$^{3,2}$
\vspace{2pt}\\
  Rachata Ausavarungnirun$^{2}$\qquad
  Gennady Pekhimenko$^{4,2}$\qquad
  Yixin Luo$^{2}$
\vspace{2pt}\\
  Onur Mutlu$^{5,2}$\qquad
  Phillip B. Gibbons$^{2,6}$\qquad
  Michael A. Kozuch$^{6}$\qquad
  Todd C. Mowry$^{2}$}
\affil{{\em $^{1}$Microsoft Research India\qquad
  $^{2}$Carnegie Mellon University\qquad
  $^{3}$NVIDIA Research}
\vspace{2pt}\\
  {\em $^{4}$University of Toronto\qquad
  $^{5}$ETH Z{\"u}rich\qquad
  $^{6}$Intel Labs}
}

\maketitle

\begin{abstract}

  This paper summarizes the idea of RowClone, 
  which was published in MICRO 2013~\cite{rowclone},
  and examines the work's significance and future potential. In existing systems, to perform
  any bulk data movement operation (copy or initialization), the data
  has to first be read into the on-chip processor, all the way into
  the L1 cache, and the result of the operation must be written back
  to main memory. This is despite the fact that these operations do
  not involve any actual computation. RowClone exploits the
  organization and operation of \ch{commodity} DRAM to perform these
  operations completely inside DRAM using two mechanisms. The first
  mechanism, Fast Parallel Mode, copies data between two rows inside
  the same DRAM subarray by issuing back-to-back activate commands to
  the source and the destination row. The second mechanism, Pipelined
  Serial Mode, transfers cache lines between two banks using the
  shared internal bus. RowClone significantly reduces the raw latency
  and energy consumption of bulk data copy and initialization. This
  reduction directly translates to improvement in performance and
  energy efficiency of systems running copy or
  initialization-intensive workloads.
  
  \ch{Our proposed technique has inspired significant research on
  various ways to perform operations in memory and reduce data 
  movement between the CPU and DRAM~\cite{lisa, li.micro17, compute-cache, kang.icassp14,
  pinatubo, shafiee.isca16, ambit,ambit-cal, kim-apbc2018, GS-DRAM}.}

\end{abstract}


\section{Problem: Bulk Data Movement}
\label{sec:problem}

The main memory subsystem is an increasingly more significant
limiter of system performance and energy efficiency\ch{~\cite{mutlu.imw13, mutlu.superfri15}} for at least
two reasons. First, the available memory bandwidth between the
processor and main memory is not growing \ch{and nor is it} expected to grow
commensurately with the compute bandwidth available in modern
multi-core processors~\cite{rlmc,jedec-bandwidth}. Second, a
significant fraction (20\% to 42\%) of the energy required to
access data from memory is consumed in driving the high-speed bus
connecting the processor and memory~\cite{dram-energy} (calculated
using \cite{micron-power}). Therefore, judicious use of the
available memory bandwidth is critical to ensure both high system
performance and energy efficiency.

In this work, we focus our attention on optimizing two important
classes of bandwidth-intensive memory operations that frequently occur
in modern systems: 1)~\emph{bulk data copy}---copying a large quantity
of data from one location in physical memory to another, and
2)~\emph{bulk data initialization}---initializing a large quantity of
data to a specific value. We refer to these two operations as
\emph{bulk data movement operations}. Prior
research\ch{~\cite{os-hardware,arch-os, kanev-isca2015}} has shown that operating systems
\ch{and data center workloads}
spend a significant portion of their time performing bulk data
movement operations. Therefore, accelerating these operations will
likely improve system performance. In fact, the x86 ISA has recently
introduced instructions to provide enhanced performance for bulk copy
and initialization (ERMSB~\cite{x86}), highlighting the importance of
bulk operations.

The main reason bulk data movement operations degrade system
performance and energy efficiency is that they require large amounts
of data to be transferred back and forth on the memory bus. This large
data transfer has three shortcomings. First, because the data \ch{is}
transferred one cache line at a time across the bus, these operations
incur high latency, directly degrading the performance of the
application performing the operation. Second, transferring a large
amount of data on the bus interferes with the memory accesses of other
concurrently-running applications, degrading their performance as
well. Finally, the large data transfer contributes to a significant
fraction of the energy consumed by these bulk movement operations.

While bulk data movement operations also degrade performance by
hogging the CPU and potentially polluting the on-chip caches, prior
works\ch{~\cite{copy-engine,bulk-copy-initialize}} have proposed simple solutions to address these
problems \ch{by adding
support for such operations in the memory controller}. However, the
techniques proposed by these works do \ch{\emph{not}} eliminate the need to
transfer data over the memory bus, which is a increasingly more
critical bottleneck for performance in modern systems.

\section{RowClone: Fast In-DRAM Copy}
\label{sec:rowclone}

The fact that both bulk data copy and initialization do \ch{\emph{not}} require
any computation on the part of the processor enables the opportunity
to perform these operations \ch{\emph{completely} inside} DRAM. Our MICRO 2013
paper~\cite{rowclone} presents a \ch{new} mechanism, RowClone, which exploits
the internal organization and operation of DRAM to perform bulk data
copy/initialization quickly and efficiently \ch{inside} DRAM.

\begin{figure}[h]
  \centering
  \includegraphics[scale=0.75]{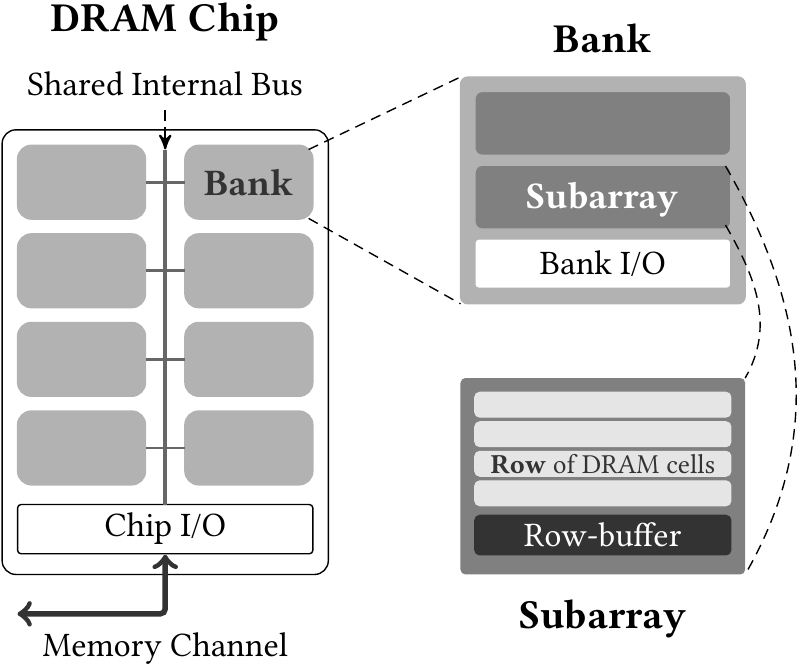}
  \caption{DRAM chip microarchitecture. Reproduced from~\cite{rowclone}.}
  \label{fig:dram-chip}
\end{figure}

Figure~\ref{fig:dram-chip} illustrates the organization of a DRAM chip.
The chip contains multiple banks, each of which is divided into subarrays,
and each subarray in turn consists of multiple rows of DRAM cells. Each
subarray contains a \emph{row buffer}, which is
used to extract the data from the DRAM cells. Data transfer
between the DRAM cells and the row buffer happen at a row
granularity, i.e., even to read a single byte from a row, the
chip copies the entire row of data from the DRAM cells to the
corresponding row buffer.\footnote{We refer the reader to our prior works\ch{~\cite{atlas,tcm,salp,tldram,al-dram,chargecache,raidr,dsarp,lisa,rowclone,chang-sigmetrics17,diva-dram,ambit,liu-isca2013, chang-sigmetrics16, hassan.hpca17, kim.cal15, lee-pact2015, lee-taco2016, kim-isca2014, patel.isca17, kim.hpca18}} for a detailed background on DRAM.}

\subsection{RowClone Mechanisms}
\label{sec:rowclone:mech}

RowClone consists of two mechanisms:
(1)~\emph{Fast Parallel Mode} (FPM), which is used to
copy data from one row to another row in the \emph{same} subarray; and
(2)~\emph{Pipelined Serial Mode} (PSM), which is used to
copy data from one row to another row in a \emph{different} subarray or bank.
We briefly discuss how each mechanism performs bulk data copy and bulk data initialization.
Section~3 of our MICRO 2013 paper~\cite{rowclone}
\ch{provides} a detailed implementation and discussion of FPM and PSM.

\textbf{Fast Parallel Mode (FPM).}
FPM uses the high internal
bandwidth offered by DRAM to quickly and efficiently copy data
between two rows within the same subarray in two simple
steps. First, FPM copies the data from the source row to the local row buffer of the subarray.
Second, FPM copies the \ch{data from} the row buffer to the
destination row. 
To perform the copy, FPM simply issues two
back-to-back \cmdact commands to the bank, first with the source
row address and the second with the destination row
address. Implementing this in existing DRAM chips requires almost
negligible changes.  \ch{These small changes are} to the peripheral logic that controls
back-to-back {\cmdact}s. 

FPM imposes two constraints on the copy operation. First, it
requires the source and the destination row to be within the same
subarray. Second, it copies the entire row's worth of data. It
cannot partially copy data from one row to another. Despite these
constraints, FPM can be used to accelerate many operations in
modern systems (Section~\ref{sec:applications}).

\textbf{Pipelined Serial Mode (PSM).}
PSM accelerates copy operations between rows in \ch{\emph{different}}
banks/subarrays,
As shown in
Figure~\ref{fig:dram-chip}, each DRAM chip uses a shared internal
bus to transfer data between the bank and the memory channel (for
both reads and writes). PSM exploits this fact to overlap the
latency of the read and write operations involved in a copy. To
implement PSM, we propose a new \ch{DRAM} command called \cmdtr. \cmdtr is
equivalent to appropriately overlapping \cmdrd to the source bank
and \cmdwr to the destination bank. However, unlike \cmdrd or
\cmdwr, \cmdtr does not transfer the data on to the memory
channel, saving significant amounts of energy. 


\textbf{Bulk Data Initialization.} For bulk initialization, RowClone
initializes one row of the destination with the required
data and then initializes the remaining rows by copying the data
from the pre-initialized row using the appropriate bulk copy
mechanism described above. For bulk zeroing (which happens
frequently), our mechanism reserves a single row in each subarray,
which is pre-initialized to zero. This enables the memory
controller to use FPM to zero out any row in the system. We refer
the reader to Section~3.4 of our MICRO 2013 paper~\cite{rowclone} for more
details on performing bulk data initialization with RowClone.

\subsection{Latency and Energy Benefits}

Table~\ref{tab:latency-energy} shows the reduction in latency and
energy consumption due to our mechanisms for different cases of 4KB
copy and zeroing operations.  To be fair to the baseline, the results
include only the energy consumed by the DRAM and the DRAM channel.  We
draw two conclusions from our results.

\begin{table}[h!]\small
  \centering
  \vspace{5pt}
 \caption{DRAM latency and memory energy reductions due to RowClone.  Adapted from~\cite{rowclone}.}
  \label{tab:latency-energy}
    \begin{tabular}{crrrrr}
  \toprule
  & \multirow{3}{*}{\textbf{Mechanism}} &
  \multicolumn{2}{c}{\small{Latency}}  & \multicolumn{2}{c}{\small{Memory Energy}}\\
  & & (ns) & ($\downarrow$) & ($\mu$J) & ($\downarrow$)\\
  \midrule
  \multirow{4}{*}{\rotatebox{90}{\textbf{Copy}}} & {Baseline} & 1046 & 1.0x & 3.6 & 1.0x\\
  & {FPM} & 90  & \textbf{11.6x} & 0.04 &  \textbf{74.4x}\\
  & {Inter-Bank - PSM} & 540  & 1.9x & 1.1 &  3.2x\\
  & {Intra-Bank - PSM} & 1050  & 1.0x & 2.5 &  1.5x\\
  \toprule
  \multirow{2}{*}{\rotatebox{90}{\textbf{Zero}}} & {Baseline} & 546  & 1.0x & 2.0  & 1.0x\\
  & {FPM} & 90  & \textbf{6.0x} & 0.05 &  \textbf{41.5x}\\
  \bottomrule
\end{tabular}

 \end{table}

First, FPM significantly improves both the latency and the energy
consumed by bulk data operations --- 11.6x and 6x reduction in latency
of 4KB copy and zeroing, and 74.4x and 41.5x reduction in memory
energy of 4KB copy and zeroing. Second, although PSM does not provide
as much benefit as FPM, it still reduces the latency and energy of a
4KB inter-bank copy by 1.9x and 3.2x, while providing a more generally
applicable mechanism.
\ch{As we show in Section~\ref{sec:results}, these latency and energy
benefits translate to significant improvements in both overall
system performance and energy efficiency.}


\subsection{End-to-End System Design}
\label{sec:system-design}
\label{sec:rowclone:zi}

To fully extract the potential benefits of RowClone, changes are
required to the ISA, processor microarchitecture, and the system
software. First, we introduce two new instructions to the ISA, namely,
\mcpy and \minit, which enable the software to indicate occurrences of
bulk data operations to the processor. Second, for each instance of
the \mcpy/\minit instruction, the processor microarchitecture
determines if the operation can be partially/fully accelerated by
RowClone and issues appropriate commands to the memory
controller. While existing mechanisms to handle Direct Memory Access
requests can be used to ensure cache coherence with RowClone, we also
propose two simple mechanisms, called \emph{in-cache copy} and \emph{clean zero
  cache line insertion}, to further reduce memory bandwidth
requirements and improve performance. We call this optimized
version of RowClone, which includes in-cache copy and clean zero cache
line insertion, \emph{RowClone-ZI}. 
Third, to maximize the use
of FPM, we make the system software aware of subarrays and the minimum
granularity of copy (required by FPM). Section~4 of our MICRO 2013
paper~\cite{rowclone} describes these changes in detail.

\section{Applications}
\label{sec:applications}

RowClone can be used to accelerate any bulk copy and
initialization operation to improve both system performance and
energy efficiency. We quantitatively evaluate the
efficacy of RowClone by using it to accelerate two primitives
widely used by modern system software: 1)~Copy-on-Write and
2)~Bulk Zeroing. We first describe these primitives, and then
discuss several applications that frequently trigger the primitives.

\subsection{Primitives Accelerated by RowClone}
\label{sec:apps-primitives}

\emph{Copy-on-Write} (CoW) is a technique used by most modern
operating systems (OS) to postpone an expensive copy operation until
it is actually needed. When data of one virtual page needs to be
copied to another, instead of creating a copy, the OS points both
virtual pages to the same physical page (source) and marks the page as
read-only. In the future, when one of the sharers attempts to write to
the page, the OS allocates a new physical page (destination) for the
writer and copies the contents of the source page to the newly
allocated page. Fortunately, prior to allocating the destination page,
the OS already knows the location of the source physical
page. Therefore, it can ensure that the destination is allocated in
the same subarray as the source, thereby enabling the processor to use
FPM to perform the copy.

\emph{Bulk Zeroing} (BuZ) is an operation where a large block of
memory is zeroed out. Our mechanism maintains a reserved row that is
fully initialized to zero in each subarray. For each row in the
destination region to be zeroed out, the processor uses FPM to copy
the data from the reserved zero-row of the corresponding subarray to
the destination row.

\subsection{Applications That Use CoW/BuZ}
\label{sec:apps-cow-zeroing}
We now describe seven example applications or use-cases that
extensively use the CoW or BuZ operations. Note that these are just a
small number of example scenarios that incur a large number of copy
and initialization operations.
\ch{Some other applications and scenarios are provided in one of our
more recent works~\cite{seshadri-isca2015}.
Recent work from Google~\cite{kanev-isca2015} shows that a considerable fraction of
execution time is spent on \texttt{memset} and \texttt{memcpy} system calls in
Google's data center workloads.}

\textit{Process Forking.} \fork is a frequently-used system call in
modern operating systems (OS). When a process (parent) calls \fork, it
creates a new process (child) with the exact same memory image and
execution state as the parent. This semantics of \fork makes it useful
for different scenarios. Common uses of the \fork system call are to
1)~create new processes, and 2)~create stateful threads from a single
parent thread in multi-threaded programs. One main limitation of \fork
is that it results in a CoW operation whenever the child/parent
updates a shared page. Hence, despite its wide usage, as a result of
the large number of copy operations triggered by \fork, it remains one
of the most expensive system calls in terms of memory
performance~\cite{fork-exp}.

\textit{Initializing Large Data Structures.} Initializing large
data structures often triggers Bulk Zeroing. In fact, many managed
languages (e.g., C\#, Java, PHP) require zero initialization of
variables to ensure memory safety~\cite{nothing}. In such cases,
to reduce the overhead of zeroing, memory is zeroed-out in bulk.

\textit{Secure Deallocation.} Most operating systems (e.g.,
Linux~\cite{linux}, Windows~\cite{windows}, Mac OS X~\cite{macos})
zero out pages newly allocated to a process. This is done to prevent
malicious processes from gaining access to the data that previously
belonged to other processes or the kernel itself. Not doing so can
potentially lead to security vulnerabilities, as shown by prior
works~\cite{shredding,sunshine,coldboot,disclosure}.

\textit{Process Checkpointing.} Checkpointing is an operation
during which a consistent version of a process state is backed-up,
so that the process can be restored from that state in the
future. This checkpoint-restore primitive is useful in many cases
including high-performance computing servers~\cite{plfs}, software
debugging with reduced overhead~\cite{flashback}, hardware-level
fault and bug tolerance mechanisms\ch{~\cite{kypros-ace,kypros-bug,li.date08,li.iccad09,li.vts10}},
and speculative OS optimizations to improve
performance~\cite{os-speculation-2,os-speculation}. However, to
ensure that the checkpoint is consistent (i.e., the original process
does not update data while the checkpointing is in progress), the
pages of the process are marked with copy-on-write. As a result,
checkpointing often results in a large number of CoW operations. 

\textit{Virtual Machine Cloning/Deduplication.}  Virtual machine
(VM) cloning~\cite{snowflock} is a technique to significantly
reduce the startup cost of VMs in a cloud computing
server. Similarly, deduplication is a technique employed by modern
hypervisors~\cite{vmware-esx} to reduce the overall memory
capacity requirements of VMs. With this technique, different VMs
share physical pages that contain the same data. Similar to
forking, both these operations likely result in a large number of
CoW operations for pages shared across VMs\ch{~\cite{seshadri-isca2015}}.

\textit{Page Migration.} Bank conflicts, i.e., concurrent requests to
different rows within the same bank, typically result in reduced row
buffer hit rate and hence degrade both system performance and energy
efficiency\ch{~\cite{salp}}. Prior work~\cite{micropages} proposed techniques to
mitigate bank conflicts using page migration. The PSM mode of RowClone
can be used in conjunction with such techniques to 1)~significantly
reduce the migration latency and 2)~make the migrations more
energy-efficient.

\textit{CPU-GPU Communication.} In many current and future processors,
the GPU is or is expected to be integrated on the same chip with the
CPU. Even in such systems where the CPU and GPU share the same
off-chip memory, the off-chip memory is partitioned between the two
devices. As a consequence, whenever a CPU program wants to offload
some computation to the GPU, it has to copy all the necessary data
from the CPU address space to the GPU address
space~\cite{cpu-gpu}. When the GPU computation is finished, all the
data needs to be copied back to the CPU address space. This copying
involves a significant overhead. By spreading out the GPU address
space over all subarrays and mapping the application data
appropriately, RowClone can significantly speed up these copy
operations. Note that communication between different processors and
accelerators in a heterogeneous system-on-chip (SoC) is done
similarly to the CPU-GPU communication and can also be accelerated by
RowClone.

\section{Results}
\label{sec:results}

In this section, we briefly summarize our evaluation of RowClone.
We evaluate three configurations:
\emph{Baseline}, an unmodified main memory subsystem that cannot
perform bulk data copy or initialization within memory;
\emph{RowClone}, which uses the FPM and PSM mechanisms described in
Section~\ref{sec:rowclone:mech}; and
\emph{RowClone-ZI}, an optimized version of RowClone that includes
the two optimizations discussed in Section~\ref{sec:rowclone:zi}.
Section~6 of our MICRO 2013 paper~\cite{rowclone} discusses our
full evaluation methodology, including details
on the simulator, system configuration, and benchmarks used for
our evaluations.

\subsection{Single-Core Evaluations}

\begin{figure}[h]
  \centering
  \includegraphics{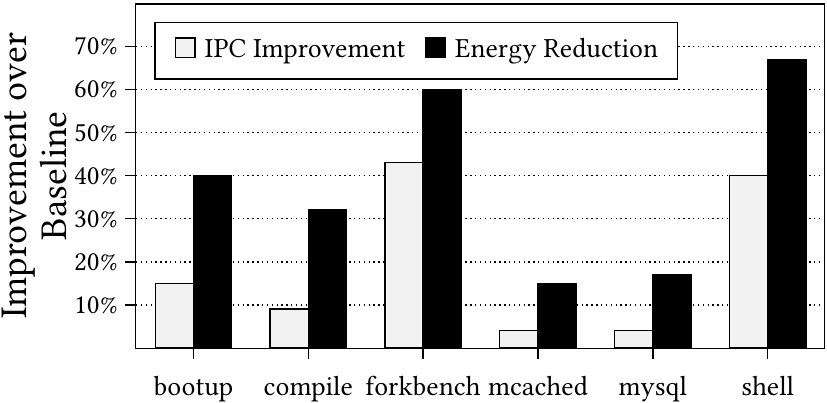}
  \caption{Performance improvement and energy reduction of RowClone-ZI 
  compared to a baseline memory subsystem without bulk copy support.}
  \label{plot:perf-energy}
\end{figure}

Figure~\ref{plot:perf-energy} shows the performance improvement and
reduction in DRAM energy consumption due to RowClone-ZI compared to the
baseline for six copy- and initialization-intensive benchmarks. As we
observe from the figure, these applications improve significantly with
RowClone-ZI.  Compared with Baseline, RowClone-ZI improves the IPC
by up to 43\%, while reducing DRAM energy consumption by up to 67\%.

Section~7 of our MICRO 2013 paper~\cite{rowclone} provides
more detailed single-core results, including (1)~the individual performance of
the FPM and PSM mechanisms using a fork benchmark (Section~7.2 of \cite{rowclone}); 
(2)~a breakdown of memory traffic for
each application into read, write, copy, and initialization operations (Section~7.3 of \cite{rowclone}); 
(3)~the performance, energy, and bandwidth improvements of both
RowClone and RowClone-ZI (Section~7.3 of \cite{rowclone}); and
(4)~a comparison of RowClone to a memory-controller-based DMA approach for
data copy and initialization, similar to~\cite{copy-engine} (Section~7.5 of \cite{rowclone}).

\subsection{Multi-Core Evaluations}
\label{sec:multi-core}

As RowClone performs bulk data operations completely within DRAM, it
significantly reduces the memory bandwidth consumed by these
operations. As a result, RowClone can benefit other applications
that are running concurrently on the same system, even if these applications
do not perform bulk data operations themselves. We evaluate this benefit of
RowClone by running our copy/initialization-intensive applications
alongside memory-intensive applications from the SPEC CPU2006
benchmark suite~\cite{spec} (i.e., those applications with last-level
cache misses per kilo-instruction, or MPKI, greater than 1). Table~\ref{tab:benchmarks} lists the set
of applications used for our multi-programmed workloads.

\begin{table}[h]\small
  \centering
  \vspace{5pt}
  \caption{List of benchmarks used for multi-core evaluation.  Reproduced from \cite{rowclone}.}
  \begin{tabular}{p{3.2in}}
  \toprule
  \textbf{Copy/Initialization-intensive benchmarks}\vspace{0.5mm}\\
  \emph{bootup},
  \emph{compile}, \emph{forkbench}, \emph{mcached}, \emph{mysql}, \emph{shell}\vspace{1mm}\\
  \toprule
  \textbf{Memory-intensive benchmarks from SPEC CPU2006}\vspace{0.5mm}\\
  \emph{bzip2}, \emph{gcc}, \emph{mcf}, \emph{milc}, \emph{zeusmp},
  \emph{gromacs}, \emph{cactusADM}, \emph{leslie3d}, \emph{namd},
  \emph{gobmk}, \emph{dealII}, \emph{soplex}, \emph{hmmer},
  \emph{sjeng}, \emph{GemsFDTD}, \emph{libquantum}, \emph{h264ref},
  \emph{lbm}, \emph{omnetpp}, \emph{astar}, \emph{wrf},
  \emph{sphinx3}, \emph{xalancbmk}\vspace{1mm}\\
  \bottomrule
\end{tabular}

  \label{tab:benchmarks}
\end{table}

We generate multi-programmed workloads for two-core, four-core and
eight-core systems. In each workload, half of the cores run
copy/initialization-intensive benchmarks, while the remaining cores
run memory-intensive SPEC benchmarks. Benchmarks from each
category are chosen at random.

\begin{figure}[h]
  \centering
  \includegraphics[scale=0.9]{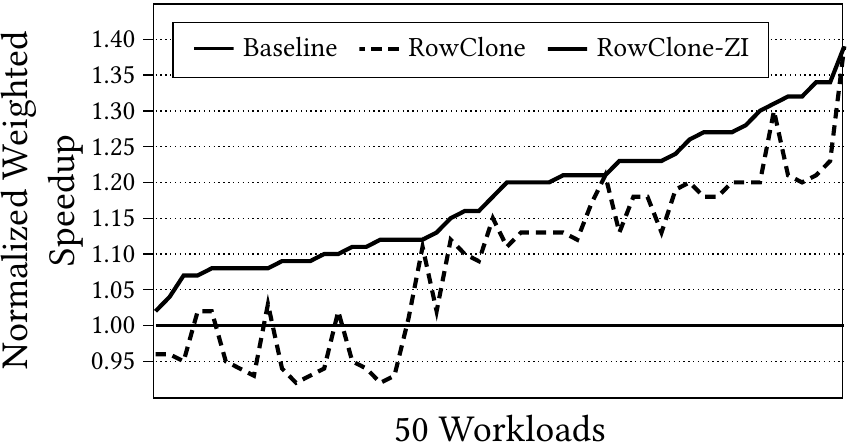}
  \caption{System performance improvement of RowClone for four-core workloads. Reproduced from~\cite{rowclone}.}
  \label{plot:s-curve}
\end{figure}

Figure~\ref{plot:s-curve} plots the performance improvement due to
RowClone and RowClone-ZI for the 50 four-core workloads that we evaluate
(sorted based on the performance improvement due to RowClone-ZI).  Two
conclusions are in order. First, although RowClone degrades
performance of certain four-core workloads (with \emph{compile},
\emph{mcached} or \emph{mysql} benchmarks), it significantly improves
performance for all other workloads (by 10\% across all
workloads). Second, RowClone-ZI eliminates the performance degradation
due to RowClone and consistently outperforms both the baseline and
RowClone for all workloads (20\% on average).

\begin{figure}[h!]
  \centering
  \includegraphics[scale=0.9]{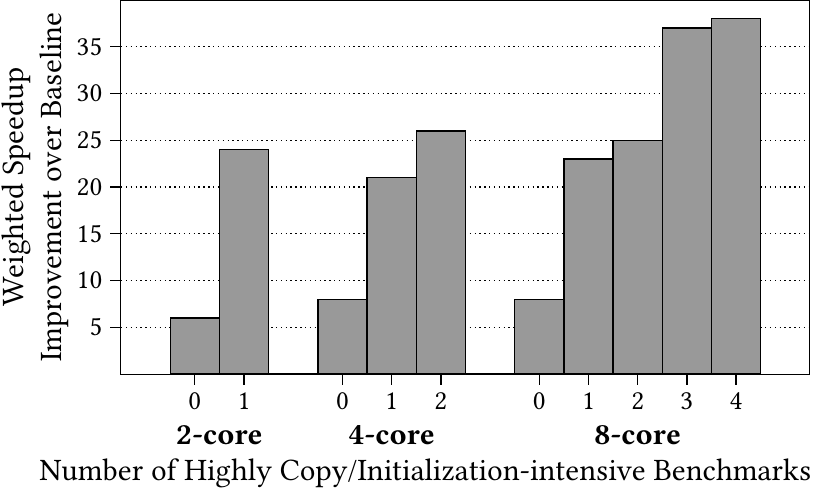}
  \caption{Effect of increasing
    copy/initialization intensity. Reproduced from~\cite{rowclone}.}
  \label{plot:multi-trend}
\end{figure}

To provide more insight into the benefits of RowClone on multi-core
systems, we classify our copy/initialization-intensive benchmarks into
two categories: 1) Moderately copy/initialization-intensive
(\emph{compile}, \emph{mcached}, and \emph{mysql}) and highly
copy/initialization-intensive (\emph{bootup}, \emph{forkbench}, and
\emph{shell}). Figure~\ref{plot:multi-trend} shows the average
improvement in weighted speedup for the different multi-core
workloads, categorized based on the number of highly
copy/initialization-intensive benchmarks. As the trends indicate, \ch{RowClone's}
performance improvement increases with increasing number of such
benchmarks for all three multi-core systems, indicating the
effectiveness of RowClone in accelerating bulk copy/initialization
operations.

We conclude that RowClone is an effective mechanism to improve system
performance, energy efficiency and bandwidth efficiency of future,
bandwidth-constrained multi-core systems.  


\section{Related Work}
\label{sec:related}

To our knowledge, this is the first paper to propose a concrete
mechanism to perform bulk data copy and initialization operations
completely in DRAM. In this section, we discuss related work and
qualitatively compare them to RowClone.
\ch{Other treatments of related works can be found in \cite{seshadri.thesis16,
seshadri.bookchapter17,seshadri.bookchapter17.arxiv}.}

\textbf{Patents on Data Copy in DRAM.} Several
patents~\cite{dram-copy-patent-1,dram-copy-patent-3,dram-copy-patent-5,dram-copy-patent-4}
propose the abstract notion that the row buffer in DRAM can be used to copy
data from one row to another. These patents have four major
drawbacks. First, they do not provide any concrete \ch{mechanism to}
perform the copy operation. Second, while using the row buffer to copy
data between two rows is possible only when the two rows are within
the same subarray, these patents make no such distinction. Third,
these patents do not discuss the support required from the other
layers of the system to realize a working system. Fourth, these
patents do not provide any concrete evaluation to show the benefits of
performing copy operations in DRAM. In contrast, RowClone is more
generally applicable, and our MICRO 2013 paper~\cite{rowclone}
discusses the concrete changes required to \emph{all
layers} of the system stack, from the DRAM architecture to the system
software, to enable bulk data copy.

\textbf{Offloading Copy/Initialization Operations.} Prior
works~\cite{bulk-copy-initialize,copy-engine} propose
mechanisms to 1)~offload bulk data copy/initialization operations to a
separate engine; 2)~reduce the impact of pipeline stalls (by
waking up instructions dependent on a copy operation as soon as
the necessary blocks are copied without waiting for the entire
copy operation to complete); and 3)~reduce cache pollution by
using hints from software to decide whether to cache blocks
involved in the copy or initialization. While Section~7.5
of our MICRO 2013 paper~\cite{rowclone}
shows the effectiveness of RowClone compared to offloading bulk
data operations to a separate engine,
techniques to reduce pipeline stalls and cache
pollution~\cite{bulk-copy-initialize} can be naturally combined
with RowClone to further improve performance.

Low-cost Interlinked Sub-Arrays (LISA)~\cite{lisa} proposes to connect
adjacent subarrays inside a DRAM bank using a set of isolation
transistors. Using this structure, LISA proposes mechanisms to
efficiently copy data across rows in different subarrays within the
same bank. LISA and RowClone can be combined to perform all bulk copy
and initialization operations efficiently inside DRAM.
However, unlike LISA, RowClone does \ch{\emph{not}} require any changes to the DRAM
array.

The Compute Cache~\cite{compute-cache} 
performs copy, zero, and bitwise operations completely inside the
on-chip SRAM cache. Like RowClone, the Compute Cache exploits the fact that many
cells are connected to the same bitline to efficiently perform these
operations across cells connected to the same bitline. Again,
depending on the location of the data, RowClone and Compute Cache can
be combined to further improve system performance and efficiency.

\textbf{Bulk Memory Initialization.}  Jarrod et
al.~\cite{heap-misses} propose a mechanism for avoiding the memory
access required to fetch uninitialized blocks on a store miss.
They use a specialized cache to keep track of uninitialized regions
of memory. RowClone can potentially be combined with this
mechanism. While Jarrod et al.'s approach can be used to reduce
bandwidth consumption for irregular initialization (initializing
different pages with different values), RowClone can be used to
push regular initialization (e.g., initializing multiple pages
with the same values) to DRAM, thereby freeing up the CPU to
perform other useful operations.


Yang et al.~\cite{nothing} propose to reduce the cost of zero
initialization by 1)~using non-temporal store instructions to
avoid cache pollution, and 2)~using idle cores/threads to perform
zeroing ahead of time. While the proposed optimizations reduce the
negative performance impact of zeroing, their mechanism does \ch{\emph{not}}
reduce memory bandwidth consumption of the bulk zeroing
operations. In contrast, RowClone significantly reduces the memory
bandwidth consumption and the associated energy overhead.

\textbf{Processing-in-Memory.}
Recent works propose mechanisms that exploit the internal organization and 
operation of DRAM~\cite{ambit,ambit-cal, li.micro17}, SRAM~\cite{compute-cache, kang.icassp14},
phase-change memory (PCM)~\cite{pinatubo}, or memristors~\cite{shafiee.isca16}
to perform bulk bitwise Boolean algebra and/or simple arithmetic operations.
One such mechanism, called Ambit~\cite{ambit,ambit-cal}, uses a number of
row copy and initialization operations to perform Boolean algebra using DRAM.
Ambit makes use of RowClone to efficiently perform these row copy and
initialization operations.
Another mechanism, the Compute Cache~\cite{compute-cache}, can perform copy and 
initialization operations within SRAM.
Other mechanisms for in-memory Boolean algebra or arithmetic~\cite{pinatubo, 
li.micro17, kang.icassp14, shafiee.isca16} can be
trivially used to perform data copy and initialization operations
\ch{(e.g., a data copy can be performed by performing a bulk addition,
where the row to be copied is added to a row of all zeroes)}.

\ch{Various} prior works
(e.g., \cite{execube,iram,ahn-isca2015a, ahn-isca2015b, zhang-hpca2014,
guo-wondp2014, boroumand-cal2016, hsieh-iccd2016, hsieh-isca2016,
pattnaik-pact2016, liu-spaa2017, kim-apbc2018, boroumand.asplos18}) have investigated mechanisms to add
logic circuitry closer to memory to perform bandwidth-intensive computations
(e.g., SIMD vector operations) more efficiently. The main
limitation of such approaches is that adding logic to or near DRAM
significantly increases the cost of main memory. In contrast, RowClone
exploits the \emph{existing} internal organization and operation of DRAM to perform
bandwidth-intensive copy and initialization operations quickly and
efficiently with low cost.

{\bf Other Methods for Lowering Memory Latency.} There are many works that
improve the performance of applications by reducing the
{\em overall memory access latency}. These works enable more
parallelism and bandwidth~\cite{salp, dsarp,
lee-taco2016, zhang-isca2014, ahn-taco2012,
ahn-cal2009, ware-iccd2006, mini-rank, ambit, ambit-cal,
lee-pact2015}, 
exploit latency variation within DRAM~\cite{al-dram, tldram, chang-sigmetrics17,
chang-sigmetrics16, diva-dram, chandrasekar-date2014},
reduce refresh counts\ch{~\cite{raidr,
liu-isca2013, khan-sigmetrics2014, khan-cal2016, venkatesan-hpca2006,
qureshi-dsn2015, khan-micro2017, khan.dsn16}}, enable better communication
between the CPU and other devices through DRAM~\cite{lee-pact2015}, leverage DRAM
access patterns \ch{to reduce access latency}~\cite{shin-hpca2014, chargecache}, reduce write-related latencies by better
designing DRAM and DRAM control policies~\cite{chatterjee-hpca2012,
lee-techreport2010, dbi}, reduce overall queuing latencies in DRAM
by better scheduling memory requests~\cite{rixner-isca2000, moscibroda-podc2008,
lee-micro2009, nesbit-micro2006, moscibroda-usenix2007,
ausavarungnirun-isca2012, ausavarungnirun-pact2015, zhao-micro2014,
mutlu-micro2007, mutlu-isca2008, ebrahimi-micro2011, atlas,
tcm, das-hpca2013, mcp, jog-sigmetrics2016,
subramanian-tpds2016, subramanian-iccd2014, rlmc, usui-taco2016,
subramanian-micro2015, mise, lee-micro2008, lee-tc2011,
li-cluster2017, pattnaik-pact2016, ghose-isca2013, kaseridis-micro2011, hur-micro2004, shao-hpca2007, mukundan-hpca2012}, employ
prefetching~\cite{fdp, patterson-sosp1995, nesbit-pact2004,
ebrahimi-hpca2009, ebrahimi-micro2009, ebrahimi-isca2011, dahlgren-tpds1995,
alameldeen-hpca2007, cao-sigmetrics1995, lee-micro2008, mutlu-hpca2003,
mutlu-ieeemicro2003, mutlu-isca2005, mutlu-micro2005, dundas-ics1997,
cooksey-asplos2002, effra}, perform memory/cache
compression~\cite{pekhimenko-micro2013, pekhimenko-pact2012, shafiee-hpca2014,
zhang-asplos2000, wilson-atec1999, dusser-ics2009, douglis-usenix1993,
decastro-sbacpad2003, alameldeen-tech2004, alameldeen-isca2004, abali-ibm2001,
pekhimenko-hpca2015, pekhimenko-hpca2016, vijaykumar-isca2015,
pekhimenko-ieeecal2015}, or perform better caching~\cite{seshadri-pact2012,
khan-hpca2014, qureshi-isca2007, qureshi-isca2006, seshadri-taco2015}.
RowClone is orthogonal to all of these approaches, and can be
combined with any of them with them to achieve higher latency and energy benefits.



\section{Significance}
\label{sec:significance}

Our MICRO 2013 paper~\cite{rowclone} proposes RowClone, a simple
mechanism to export bulk copy and initialization operations to
DRAM. In this section, we describe the novelty of our approach, the
long term impact of our proposed techniques, and new research
directions triggered by our work.

\subsection{Novelty}
\label{sec:novelty}

Prior works investigate mechanisms to add logic closer to
memory to perform bandwidth-intensive operations more
efficiently. Although this approach has the potential to be used for
a wide range of applications, it has two shortcomings. First, adding logic to DRAM
increases the cost of DRAM significantly. Second, this approach
does \ch{\emph{not}} reduce the bandwidth requirement of simple bulk
copy/initialization operations.

In contrast, our work is the first (to our knowledge) to propose
mechanisms that exploit the internal organization and operation of DRAM
to perform bandwidth-intensive copy and initialization operations
quickly and efficiently \emph{in} DRAM. The changes required by
our mechanism in the DRAM chip are limited to the peripheral logic
and are very modest, with a DRAM die area overhead of only 0.2\%. With this
small overhead, our mechanisms significantly reduce the latency,
bandwidth, and energy consumed by bulk data operations.

\subsection{Long-Term Impact}
\label{sec:long-term}

We believe four trends in current and future systems make our
proposed solutions even more relevant.  We discuss each trend,
and how RowClone can be applied in the context of the trend.

\textbf{Increasingly Limited Memory Bandwidth.}  Processor
manufactures are integrating more and more cores on a single chip,
thereby significantly increasing the compute capability of the
processing chip. However, due to (1)~the high cost associated with
increasing pin counts and (2)~limitations in DRAM scalability, 
the available memory bandwidth is not
expected to grow at the same rate\ch{~\cite{jedec-bandwidth, rlmc}}. This
makes mechanisms like RowClone, which significantly reduce the
overall memory bandwidth utilization of the system, likely even
more important in future systems.

\textbf{Increasing Use of Hardware Accelerators.} Many modern
processors already integrate the GPU on the same die as the
CPU. With emerging systems moving towards a system-on-chip (SoC)
model, many components/accelerators (called \emph{agents}) are
integrated on the same die as the CPU, and share the off-chip
memory\ch{~\cite{usui-taco2016, suleman-isca2010}}. 
To reduce the complexity of managing these agents, each
agent is given its own share of the physical address space, and
agents typically communicate with each other by copying data in
bulk across the individual device address spaces. By enabling
faster bulk data copies, we expect RowClone to significantly
reduce the communication latency between different agents without
increasing the complexity of the system.

\textbf{Increasing Use of Virtualization.} Modern systems
(especially data centers and cloud computers) are increasingly employing
virtualization to \ch{improve} the utilization, security, and availability
of systems and services. As described in our MICRO 2013 paper~\cite{rowclone}, the use of
techniques such as VM cloning and
deduplication~\cite{snowflock,vmware-esx} to reduce the memory
capacity requirements will likely increase the number of copy
operations and zeroing operations (to protect data across
VMs). RowClone can improve the performance and energy efficiency
of such systems by performing these copy/initialization operations
efficiently.

\textbf{Ease of Adoption.} Given the low implementation
complexity of RowClone, it can be easily adopted in existing
systems. RowClone is not limited only to DDR DRAMs. It can be used
with 3D-stacked DRAM technologies~\cite{loh2008stacked, lee-taco2016} 
such as the Hybrid Memory Cube~\cite{hmc.spec.1.1, hmc.spec.2.0}
and High Bandwidth Memory~\cite{jedec.hbm.spec},
which are gaining increasing interest among researchers, DRAM
manufacturers, and system designers\ch{~\cite{kim.cal15, ahn-isca2015a, ahn-isca2015b}}.

\subsection{New Research Directions}
\label{sec:research}

Our proposed approach to performing bulk data copy and
initialization in DRAM inspires several important research
directions (and hopefully many more that others
will imagine). We describe a few of them below.

One important research question that our work raises is \emph{how
  can one redesign system software (e.g., operating system, hypervisors)
  and application software to take better advantage of
  RowClone?}  Existing systems assume that copies are expensive
and hence trade off complexity for performance. However, with
RowClone, it may be possible to design simpler yet high
performance systems by rethinking software design in the presence
of \ch{very} fast bulk copy and initialization.

Our MICRO 2013 paper~\cite{rowclone} proposes low-cost mechanisms to export bulk copy and
initialization to DRAM. These are by no means the only
bandwidth-intensive operations. There are other operations
that unnecessarily move data between the main
memory and the processor, which can be optimized using low-cost mechanisms.
Therefore, another natural research
question is \emph{what other bandwidth-intensive operations can be
  exported to main memory using low-cost mechanisms?} We believe
RowClone can inspire similar mechanisms for other such operations.
For example, one of our recent works~\cite{GS-DRAM} proposes an efficient 
method to perform gather/scatter operations in DRAM.
\ch{Another of our recent works proposes mechanisms to perform bulk bitwise
operations in DRAM~\cite{ambit, ambit-cal}, building upon and taking advantage
of RowClone.}

Recently, there has been increased interest in emerging
non-volatile memory technologies (e.g., PCM~\cite{wong.procieee10, lee-isca2009, lee-ieeemicro2010, qureshi-isca2009,
yoon-taco2014,lee-cacm2010,qureshi-micro2009,yoon-iccd2012},
STT-MRAM~\cite{naeimi.itj13, ku-ispass2013,guo-isca2009,chang-hpca2013}, memristors~\cite{chua.tct71, strukov.nature08}). Given this trend, \emph{exploring the feasibility of
  extending RowClone to these new memory technologies} is a
relevant and important research direction.
For example, two recent works~\cite{pinatubo, shafiee.isca16}
use the principles discussed in RowClone to perform 
bulk Boolean algebra and arithmetic operations within emerging memories.
\ch{Similarly, exploring the idea of RowClone in other storage/memory
technologies, e.g., NAND flash memory~\cite{cai.procieee17, cai.procieee.arxiv17, cai.bookchapter.arxiv17},
is promising.}

Given that memory bandwidth is expected to become an even more scarce
resource in future systems, answers to these research questions
have the potential to \ch{greatly} mitigate bandwidth contention, and, thus,
significantly improve both the performance and
energy efficiency of these systems.

\subsection{Works Building on RowClone}

RowClone has inspired a number of followup works that propose
1)~new mechanisms to perform bulk operations inside
various memory technologies (e.g., DRAM~\cite{lisa, li.micro17},
SRAM~\cite{compute-cache, kang.icassp14}, PCM~\cite{pinatubo}, 
memristors~\cite{shafiee.isca16}), and 2)~mechanisms
that exploit RowClone to speedup other operations (e.g.,
in-DRAM \ch{bulk bitwise operations~\cite{ambit,ambit-cal, kim-apbc2018}}).
\ch{A survey of related works is provided in \cite{seshadri.bookchapter17}.}

%
%
%


One of our recent works, Ambit~\cite{ambit,ambit-cal}, proposes a
mechanism to perform bulk bitwise operations completely inside
DRAM. Ambit operations involve a number of row copy and initialization
operations. Ambit uses RowClone to perform these operations quickly
and efficiently inside DRAM. In fact, RowClone is essential for Ambit
to obtain the performance and energy efficiency improvements.
Other recent works that perform bulk bitwise Boolean algebra and/or
simple arithmetic operations\ch{~\cite{compute-cache, pinatubo, 
li.micro17, kang.icassp14, shafiee.isca16, levy.microelec14, kvatinsky.tcasii14,
kvatinsky.iccd11, kvatinsky.tvlsi14,akerib.patent15,akerib.patent14}} exploit the organization and
operation of memory arrays, akin to RowClone, and can be used to
perform bulk data copy and initialization operations.

Data movement is expected to become \ch{an even more} critical problem in future
systems. We believe RowClone can inspire \ch{other works} that
propose mechanisms to reduce data movement, thereby enabling higher
system performance and energy efficiency.

\section{Conclusion}
\label{sec:conclusion}

Our MICRO 2013 paper~\cite{rowclone} proposes RowClone,
a mechanism that performs bulk data
copy and initialization operations completely inside DRAM.
RowClone consists of two mechanisms, Fast Parallel Mode and
Pipelined Serial Mode, that are used to copy data using
existing peripheral structures within DRAM, requiring no changes to
the DRAM cell array.
By enabling efficient bulk data copy and initialization, 
RowClone provides \ch{significant} performance and DRAM energy 
improvements \ch{that are between one to two orders of magnitude
higher compared to existing systems}.

RowClone is one of the first steps
towards reducing unnecessary data movement between the processor
and the main memory using a low-cost in-memory approach. Current
trends in system design indicate that our approach will be more
relevant to future, bandwidth-limited systems. We hope that our
work triggers research that leads to \ch{1)~}simpler and more efficient
software design and \ch{2)~}extensions of our approach to other
operations and memory technologies, with the goal of continuing to
\ch{greatly} improve system performance and energy efficiency.

\section*{Acknowledgments}

\ch{We thank Saugata Ghose for his dedicated effort in the preparation of
this article.}
We acknowledge the support of AMD, IBM, Intel, Oracle, Qualcomm, and
Samsung. This research was partially supported by the NSF (grants 0953246,
1147397, and 1212962), the Intel University Research Office Memory
Hierarchy Program, the Intel Science and Technology Center for Cloud
Computing, and the Semiconductor Research Corporation.


\bibliographystyle{IEEEtranS}
\bibliography{references}

\end{document}